\documentstyle[epsfig]{mn}

\title{Formation of gas disks in merging galaxies}

\author[J.E. Barnes]{Joshua E. Barnes
\thanks{E-mail: barnes@ifa.hawaii.edu}\\
Institute for Astronomy, 2680 Woodlawn Dr.,
Honolulu, HI 96822, USA}

\begin{document}

\maketitle

\begin{abstract}
Observations indicate that much of the interstellar gas in merging
galaxies may settle into extended gaseous disks.  Here, I present
simulations of disk formation in mergers of gas-rich galaxies.  Up to
half of the total gas settles into embedded disks; the most massive
instances result from encounters in which both galaxies are inclined
to the orbital plane.  These disks are often warped, many have rather
complex kinematics, and roughly a quarter have counter-rotating or
otherwise decoupled central components.  Disks typically grow from the
inside out; infall from tidal tails may continue disk formation over
long periods of time.
\end{abstract}


\section{INTRODUCTION}

Gas falling into the nuclei of interacting and merging galaxies has
attracted much attention.  Nuclear inflows fuel the bursts of star
formation responsible for the anomalous colors (Toomre \& Toomre 1972;
Larson \& Tinsley 1978) and extraordinary IR luminosities (Joseph \&
Wright 1985; Sanders \& Mirabel 1996) of merging galaxies; the stars
formed in this manner may play a crucial role in transforming merger
remnants into elliptical galaxies (Kormendy \& Sanders 1992).
Numerical simulations show that these inflows occur when gravitational
torques remove angular momentum from shocked galactic gas (Noguchi
1988; Hernquist 1989; Combes, Dupraz, \& Gerin 1990; Barnes \&
Hernquist 1991, 1996).  In typical cases $\sim 60$ percent of the gas
initially distributed throughout the progenitor disks may wind up in a
nuclear cloud with dimensions of $\sim 0.1 {\rm\,kpc}$.

Less attention has been paid to the gas which {\it fails\/} to reach
the nuclei of merger remnants.  Extended H{\small I}, H{\small II},
and X-ray emission is often seen in merging galaxies, but the gas
responsible is usually too dilute to directly fuel violent star
formation or AGN activity.  There has been little numerical work on
the fate of extended gas in mergers, in part because it's expensive to
continue a simulation once a compact nuclear gas cloud has formed.

There are reasons to think that extra-nuclear gas may play a role in
the evolution of merger remnants.  The well-studied merger remnant
NGC~7252 exhibits extended emission from ionized gas with remarkably
complex kinematics (Schweizer 1982, 1998): an inner disk with radius
$\sim 3 {\rm\,kpc}$ and a well-defined axis of rotation is surrounded
by a region which shows rapid minor-axis rotation and an apparent
velocity reversal.  This inner disk, seen as a vivid spiral in HST
images (Whitmore et al. 1993), contains $\sim 10^{9.5} {\rm\,M_\odot}$
of molecular and ionized gas (Dupraz et al. 1990).  A comparable
amount of atomic gas lingers in the tidal tails; these return gas to
the main body of the remnant at at rate of $\sim 4
{\rm\,M_\odot\,yr^{-1}}$ (Hibbard et al. 1994; Hibbard \& Mihos 1995).
Whatever the fate of this gas, it's worth asking how a system with
such complex kinematics could form in the first place.

In this paper I use simulations to explore the possibility that
extended gas disks and associated structures can form as part of the
merger process.  Each simulation begins with a pair of disk galaxies
falling together on a parabolic relative orbit, and follows their
collision and the eventual formation and initial relaxation of a
merger remnant.  The interstellar material is modeled by including a
component obeying the equations of motion of an isothermal gas.  This
approach is deliberately simple-minded when compared to simulations
which include star formation, feedback, and other processes (e.g.~Katz
1992; Mihos \& Hernquist 1994; Gerritsen \& Icke 1997; Springel 2000).
Star formation is clearly important for the evolution and eventual
fate of merging galaxies, but there's no evidence that stellar
processes are essential for the formation of {\it gas\/} disks like
the one in NGC~7252.  My intention is to present a simple model which
appears capable of explaining the disks observed in merger remnants.

\section{SIMULATIONS}

Isothermal simulations are relatively cheap, so an encounter survey
can be performed with modest computational resources.  For ease of
comparison, I adopted the same set of four encounter geometries used
in previous studies (e.g.~Barnes 1992; Barnes \& Hernquist 1996).
Table~\ref{tab:geometries} lists disk inclinations $i$ and pericentric
arguments $\omega$ for the passages used in this paper.  Each
combination was used in three different encounters: first, a close
passage of galaxies with a 1:1 mass ratio; second, a more distant
passage, again with a 1:1 ratio; and third, a close passage with a 3:1
mass ratio.  Thus there are $12$ simulations in this survey.  For
convenience, I designate each simulation by specifying the disk
geometry (from Table~\ref{tab:geometries}), the mass ratio (either 1:1
or 3:1), and optionally the pericentric separation (C or D).  Thus
DIR~1:1~C specifies a direct, close passage of equal-mass galaxies,
DIR~1:1~D is a more distant version of the same encounter, and INC~3:1
is an inclined passage of galaxies with a 3:1 mass ratio.

\begin{table}
\caption{Disk angles for encounter survey.}
\label{tab:geometries}
\begin{tabular}{@{}lrrrr@{}}
{\bf Geometry} & $i_1$ & $\omega_1$ & $i_2$ & $\omega_2$ \\
\\
DIRect     &     0 &         -- &    71 &         30 \\
RETrograde &   180 &         -- &  -109 &         30 \\
POLar      &    71 &         90 &  -109 &         90 \\
INClined   &    71 &        -30 &  -109 &        -30 \\
\end{tabular}
\end{table}

The galaxy models in these experiments are the same as those used in
earlier studies (Barnes 1998; Bendo \& Barnes 2000).  Briefly, each
model contains a bulge with a shallow cusp (Hernquist 1990), an
exponential disk with constant scale height (Freeman 1970; Spitzer
1942), and a dark halo with a constant-density core (Dehnen 1993;
Tremaine et al.~1994).  Density profiles for these components are
\begin{equation}
  \begin{array}{@{}l}
    \rho_{\rm bulge} \propto
      r^{-1} (r + a_{\rm bulge})^{-3} \,, \\
    \rho_{\rm disk} \propto
      \exp(-R/R_{\rm disk}) \, {\rm sech}^2(z/z_{\rm disk}) \,, \\
    \rho_{\rm halo} \propto
      (r + a_{\rm halo})^{-4} \,,
  \end{array}
\end{equation}
where $r$ is the spherical radius, $R$ is the cylindrical radius in
the disk plane, and $z$ is the distance from the disk plane.  The gas
is initially distributed like the disk component and amounts to $12.5$
percent of the disk mass.  Initial velocities and velocity dispersions
for the collisionless components were determined from the Jeans
equations (e.g.~Hernquist 1993); the disk gas was set rotating at the
local circular velocity.

Unless otherwise indicated, the results below are given in simulation
units with $G = 1$.  In these units, the galaxies in the equal-mass
encounters had total masses $M_1 = M_2 = M_{\rm bulge} \! + \! M_{\rm
disk} \! + \!  M_{\rm halo} = \frac{1}{16} \! + \! \frac{3}{16} \! +
\! 1 = \frac{5}{4}$ and length scales $a_{\rm bulge} = 0.04168$,
$R_{\rm disk} = 0.08333$, $z_{\rm disk} = 0.007$, and $a_{\rm halo} =
0.1$.  These choices yield a model with half-mass radius $r_{\rm half}
\simeq 0.28$, rotation period $t_{\rm rot}(r_{\rm half}) \simeq 1.2$,
and binding energy $E \simeq -1.07$.  The same model was used for the
larger galaxy in the simulations with a 3:1 mass ratio, while the
smaller model was simply scaled down by a factor of $3$ in mass and a
factor of $\sqrt{3}$ in radius, thereby following a $M \propto v^4$
relation similar to the observed luminosity-rotation velocity relation
for disk galaxies.  The large galaxy model may be roughly scaled to
the Milky Way by equating the simulation units of length, mass, and
time to $40 {\rm\,kpc}$, $2.2 \times 10^{11} {\rm\,M_\odot}$, and $2.5
\times 10^8 {\rm\,yr}$, respectively.  Gravitational forces were
calculated after smoothing the mass distribution using a Plummer
kernel with $\epsilon = 0.0125$.

The gas had an isothermal equation of state, $P = c_{\rm s}^2 \rho$,
where the sound speed is fixed at $c_{\rm s} = 0.0966$ velocity units.
The simulation code evaluates the heating $\dot{u}$ of each gas
particle due to adiabatic compression and shocks; this energy is
assumed to be radiated instantly, so the gas stays at a constant
temperature.  The sound speed $c_{\rm s}$ is an order of magnitude
smaller than typical circular velocities in these models; thus gas
pressure forces are relatively small, and the gas travels on roughly
ballistic trajectories except where diverted by shocks.  If the large
galaxy model is scaled to the MW, the sound speed is $c_{\rm s} \simeq
15.5 {\rm\,km\,s^{-1}}$, corresponding to a gas temperature of $\sim
20000 {\rm\,K}$.  I chose this value to take account of non-thermal
pressures due to magnetic fields and turbulence, but the more
conventional choice of $10^4 {\rm\,K}$ would yield nearly identical
results.

Initial conditions for the encounters were generated by building pairs
of galaxy models, placing them on the chosen orbits, and rotating them
to the desired orientations.  The initial center-of-mass position and
velocity of each galaxy were determined from a parabolic orbit of two
point masses $M_1$ and $M_2$ with pericentric separation $r_{\rm
peri}$ and time of pericenter $t_{\rm peri}$.  For the 1:1 encounters
the close passages had $r_{\rm peri} = 0.2$, while the distant passage
had $r_{\rm peri} = 0.4$.  The 3:1 encounters used $r_{\rm peri} =
0.2$.  In all cases $t_{\rm peri} = 1$, so point masses on these
orbits would reach pericenter exactly one time unit after the start of
the simulation.  Of course, the actual trajectories deviate from the
Keplerian ideal as soon as the galaxy models begin to interpenetrate
(e.g.~Barnes 1988, 1992).  At {\it first\/} passage, these deviations
are modest -- the close encounters come within $r_{\rm peri} \simeq
0.25$ at $t_{\rm peri} \simeq 1.03$, while the distant ones reach
$r_{\rm peri} \simeq 0.45$ at $t_{\rm peri} \simeq 1.04$.

Each simulation used $N_{\rm gas} \! + \! N_{\rm star} \! + \! N_{\rm
halo} = 24576 \! + \!  29696 \! + \! 32768 = 87040$ particles.
Collisionless components were followed using standard N-body
techniques, while the gas was simulated using Smoothed Particle
Hydrodynamics or SPH (e.g.~Monaghan 1992); the code uses a
hierarchical algorithm to compute gravitational forces, adaptive
smoothing to resolve a fixed mass scale in the gas, and adaptive
time-steps determined by a Courant condition.  To take maximum
advantage of a cluster of eight $333 {\rm\,MHz}$ processors, I ran
multiple calculations simultaneously.  Most runs required between
$400$ and $800$ processor hours, largely spent in the later stages
where high gas densities demand very short time-steps.  Energy and
angular momentum were conserved to a fraction of a percent in all
cases.

\begin{figure*}
\begin{center}
\epsfig{figure=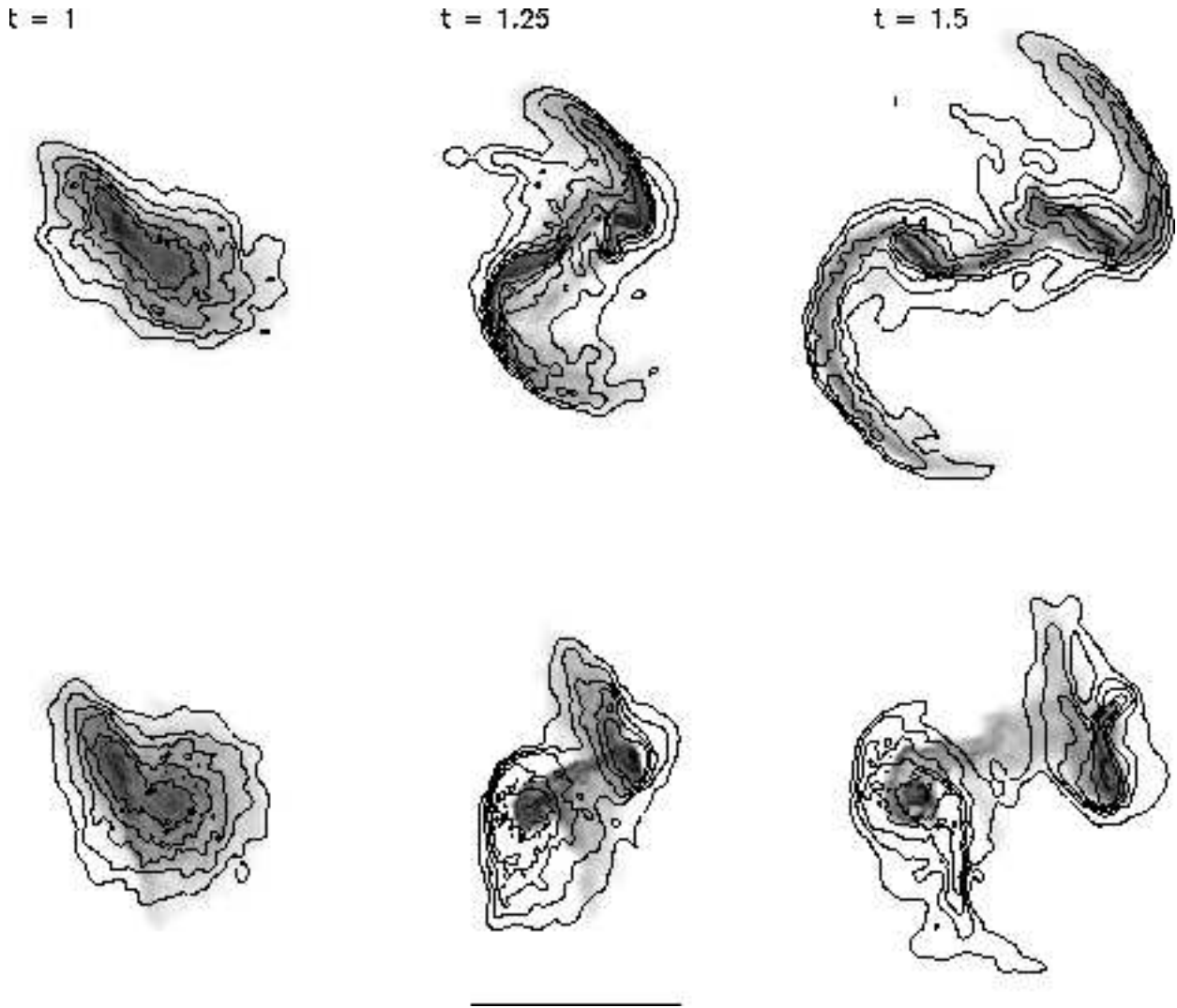,width=6.5in}
\end{center}
\caption{First passages of DIR~1:1~C (above) and RET~1:1~C (below),
each viewed along the orbital axis.  Grey-scale shows logarithmic gas
surface density; adaptive smoothing is used to preserve details in
high-density regions.  Contours show stellar surface density in steps
of one magnitude; lowest contour is $\Sigma = 0.03125$.  The scale bar
at bottom is one length unit long.  Times are given at the top.}
\label{fig01}
\end{figure*}

The resolution of these simulations is comparable to that available in
earlier SPH studies of dissipative encounters involving disk galaxies
(e.g.~Hernquist 1989; Barnes \& Hernquist 1991, 1996; Mihos \&
Hernquist 1994, 1996).  Gravitational forces were evaluated with a
fixed smoothing of $\epsilon = 0.0125$ length units, which is small
enough to follow the overall dynamics of the interacting galaxies; the
nuclei of the resulting merger remnants are poorly resolved, but the
disks formed in these simulations are an order of magnitude larger
than $\epsilon$ and thus not severely compromised by gravitational
smoothing.  Hydrodynamic forces were calculated by smoothing over $40$
gas particles; each gas particle has a mass of $2^{-19} \simeq 1.9
\times 10^{-6}$ mass units, so the SPH calculation has a mass
resolution of $7.6 \times 10^{-5}$.  For gas with a density $\rho$ and
sound speed $c_{\rm s} = 0.0966$, the Jeans length is $\lambda_{\rm J}
\simeq 0.1712 \rho^{-1/2}$ and the Jeans mass is $M_{\rm J} \simeq
0.00263 \rho^{-1/2}$; the gravitational force calculation resolves
$\lambda_{\rm J}$ if $\rho \la 190$, while the SPH calculation
resolves $M_{\rm J}$ if $\rho \la 1200$.  Thus the collapse of
small-scale structure in the gas is limited by the spatial resolution
of the gravitational force calculation, with hydrodynamic resolution
playing only a secondary role.  Suppression of small-scale collapse,
which would be fatal for studies of fragmentation via Jeans
instabilities (e.g.~Bate \& Burkert 1997), may even benefit these
simulations by preserving the smooth structure of the gas.  The code
does {\it not\/} suppress the Jeans instability on scales
$\lambda_{\rm J} > \epsilon$; on such scales, the gas is usually
stabilized by the combined effects of pressure and rotation (Toomre
1964).

Nonetheless, these simulations are in fact quite crude.  The SPH
smoothing length often exceeds the vertical scale height of the
simulated gas disks; in effect, these structures are resolved in the
radial direction, but not in the vertical direction.  Under such
circumstances the calculation provides only a rough approximation to
the dynamics of a smooth gas, and it seems better to view the SPH code
as a locally momentum-conserving scheme in which particles
representing gas tend to seek out closed, non-intersecting orbits.  At
least an order of magnitude more gas particles are needed to
significantly improve this situation.  While there's no simple way to
anticipate the results of such ambitious calculations, it's plausible
that disk-like structures like those reported in this paper will
continue to appear in future experiments.

\section{ENCOUNTERS}


Direct and retrograde versions of a close passage are compared in
Figure~\ref{fig01}.  Encounter DIR~1:1~C (top) resembles direct
encounters described in other studies (Barnes \& Hernquist 1991, 1996;
Mihos \& Hernquist 1994, 1996).  Although the disks interpenetrate,
only a modest fraction of the gas actually collides with gas from the
other disk.  The stellar and gaseous components both respond to the
tidal forces by developing extended bridges and tails (Toomre \&
Toomre 1972).  In the aftermath of such passages, much of the gas is
rapidly driven into the central regions; these inflows are driven by
gravitational torque between the stellar and gaseous bars formed in
tidally perturbed disks (Combes, Dupraz, \& Gerin 1990; Barnes \&
Hernquist 1991; Mihos \& Hernquist 1996).  Encounter RET~1:1~C
(bottom) illustrates a violent {\it hydrodynamic\/} interaction.  The
geometry of this collision insures that a large fraction of the gas
suffers strong shocks as the galaxies intersect.  By time $t = 1.25$
(middle), much of the gas in these disks has been swept toward their
centers.  These inflows are not driven by gravitational torques -- the
stellar disks are far less perturbed than the gas, and incapable of
exerting strong gravitational torques.  Instead, they are driven by
hydrodynamic forces; the gas loses its spin angular momentum by
colliding directly with gas in the other galaxy.  Gas which escapes
being swept in forms a plume connecting the two galaxies; such
structures may have been observed in some deeply interpenetrating
encounters (e.g.~Condon et al.~1993; Tsuchiya, Korchagin, \& Wada
1998).

\begin{figure}
\begin{center}
\epsfig{figure=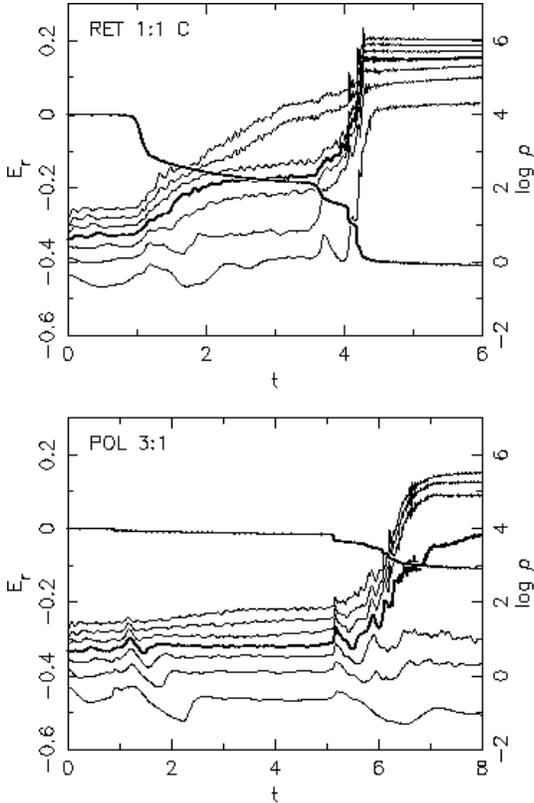,width=2.75in}
\end{center}
\caption{Thermodynamic indicators plotted as functions of time $t$.
Falling curves show $E_{\rm rad}$, the energy lost to dissipation.
Rising curves show the first through seventh octiles of the gas
density, $\rho$; the heavier line indicates the median density.}
\label{fig02}
\end{figure}

\begin{figure*}
\begin{center}
\epsfig{figure=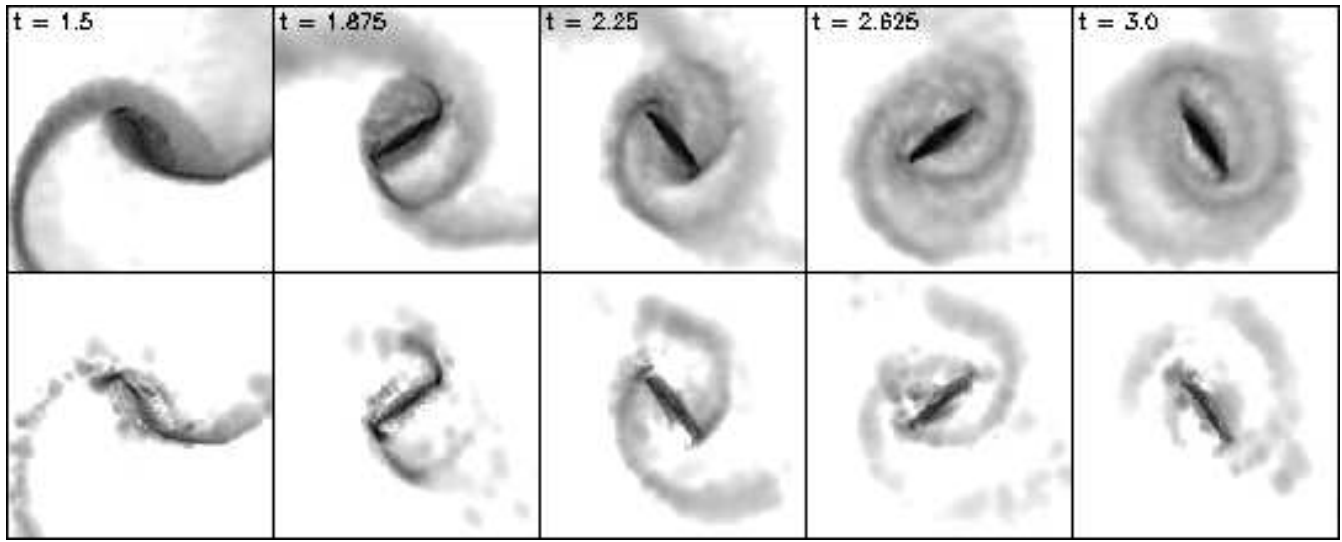,width=\textwidth}
\end{center}
\caption{Evolution of large disk in DIR~3:1 after first passage.  The
upper frames weights all gas particles equally; the lower frames show
gas weighted by energy dissipation.  All frames are $0.8 \times 0.8$
length units.}
\label{fig03}
\end{figure*}

To illustrate the range of evolutionary histories in this sample of
encounters, Figure~\ref{fig02} concisely summarizes two rather
different experiments.  In these plots, $E_{\rm rad}$ is the total
energy lost to radiative processes, while $\rho$ is the gas density.
As one might guess from the discussion above, encounter RET~1:1~C is
the most dissipative of those studied here; net radiative losses
amount to $\sim 20$ percent of the initial binding energy of the
entire system.  These losses occur in large-scale shocks as the two
galaxies plow into each other at $t \simeq 1$, fall back together at
$t \simeq 3.7$, and merge at $t \simeq 4.3$.  Gas densities increase
with each burst of dissipation; by the end of the simulation $\sim 90$
percent of the gas lies in a barely-resolved disk at the center of the
merger remnant.  In contrast, encounter POL~3:1 is the least
dissipative, losing only $\sim 5$ percent of its initial binding
energy.  The first passage at $t \simeq 1$, while close enough to
produce definite tidal features, barely registers in the traces of
$E_{\rm rad}$ and $\rho$.  The next passage, at $t \simeq 5.0$, is
more dramatic, and the final merger at $t \simeq 6.3$ drives about
half of the gas into the central regions; the rest of the gas
eventually settles into a warped disk.

Disk formation need not wait until the merger process is completed;
reaccretion from tidal bridges and tails can feed high angular
momentum gas back into galaxies after any reasonably direct passage.
An example is illustrated in Figure~\ref{fig03}.  Here the upper row
shows gas in the larger disk of encounter DIR~3:1 responding after the
direct passage of its lighter companion; note the pronounced bar
typically formed in such passages.  The lower row shows where shocks
occur by rendering each particle with intensity proportional to the
local dissipation rate $\dot{u}$.  While the central bar is prominent
in the lower images, shocks also develop in the disk surrounding the
bar.  Some of these shocks form where gas falling back from the tidal
tail and bridge impinges on the disk; others occur within the disk,
perhaps in response to forcing by the central bar.  Between times $t =
1.875$ and $t = 3.0$ the reaccreted gas nearly doubles the size of the
disk.


Reaccretion of gas from tidal tails is evident in a recent H{\small I}
study of NGC~4038/9 (Hibbard et al.~2001).  Relative to the systemic
velocity, most of the tail associated with the northern disk
(NGC~4038) is moving away from us, but the gas at the base of the tail
has the opposite sense of motion.  This indicates that the material at
the base has already attained apocenter and is now falling back onto
the disk.  This reaccreted gas may be fueling the ring of star
formation in the disk of NGC~4038.

\begin{figure*}
\begin{center}
\epsfig{figure=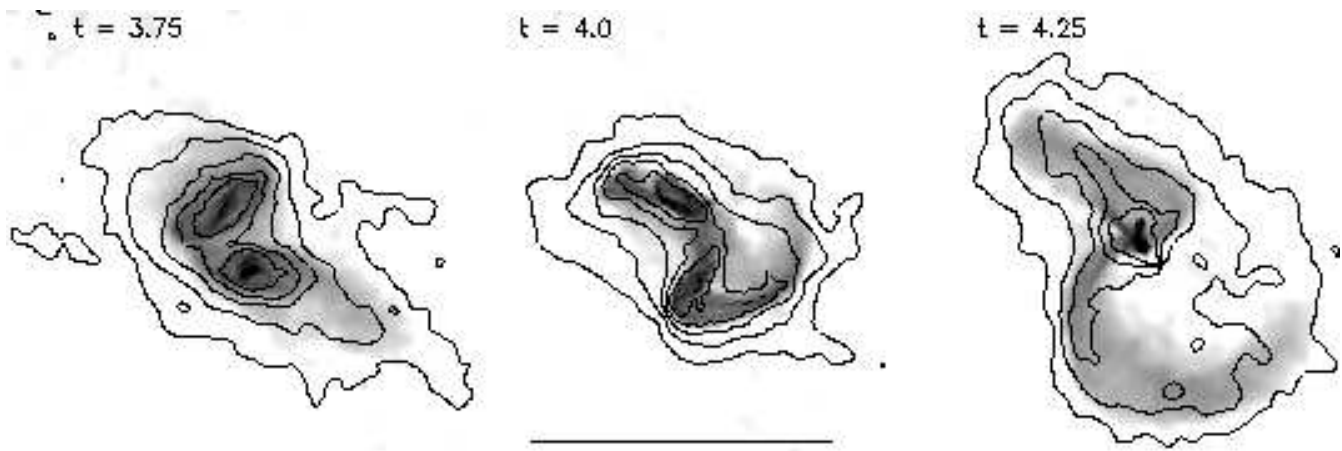,width=\textwidth}
\end{center}
\caption{Second passage of POL~1:1~C.  Grey-scale, contours, scale bar
as in Fig.~\ref{fig01}.}
\label{fig04}
\end{figure*}

One consequence of this disk rebuilding is that the gas and stars may
be kinematically segregated in tidally disturbed disks.  Reaccreted
gas seeks out closed orbits, usually close to the disk plane, while
reaccreted stars remain on highly eccentric orbits.  This can produce
dramatic effects in further tidal interactions, possibly explaining
some cases in which stellar and H{\small I} tails fail to coincide
(Hibbard, Vacca, \& Yun 2000, Mihos 2001).  Figure~\ref{fig04}
illustrates the behavior of gas and stars in the {\it second\/}
passage of encounter POL~1:1~C.  The first frame shows two fairly
relaxed gas disks (grey-scale), each embedded in a distended stellar
disk (contours).  In the second frame the disks have moved past each
other, and a pronounced tail extends from the more direct member of
the pair.  By the third frame the gas and stellar distributions are
quite different -- the one is no longer a good predictor of the other.
For example, the gas in the lower tail crosses several contours of the
stellar distribution; evidently the kinematically cold gas has
produced a narrow tidal tail, while the warmer stellar component has
raised a broader feature.  This last frame also shows that the two
galaxies are at the point of merging.

\section{REMNANTS}

\begin{figure*}
\begin{center}
\epsfig{figure=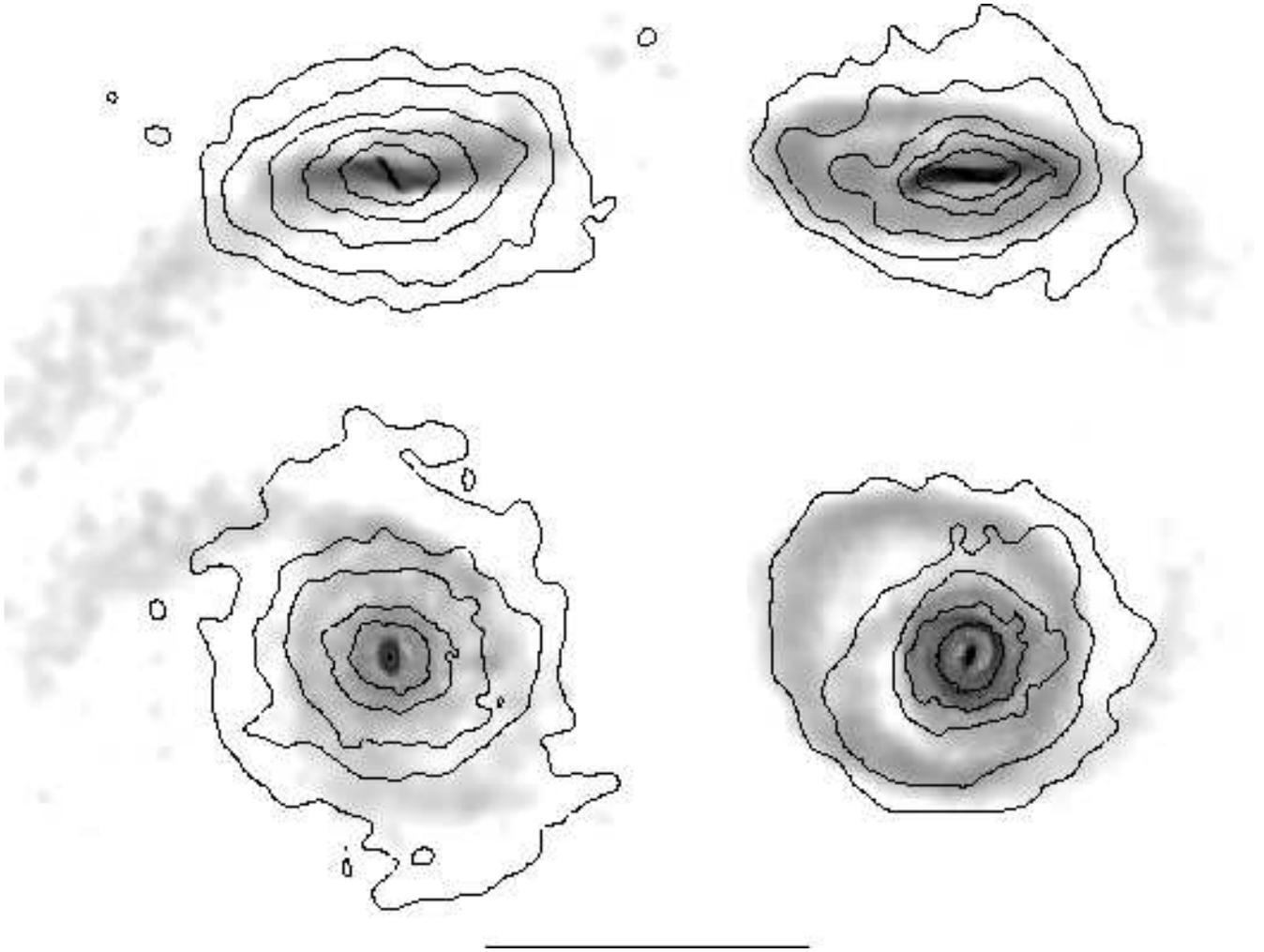,width=\textwidth}
\end{center}
\caption{Edge-on (top) and face-on (bottom) views of remnants produced
by encounters DIR 1:1 C at $t = 6$ (left) and POL 3:1 at $t = 8$
(right).  Grey-scale, contours, scale bar as in Fig.~\ref{fig01}.  The
stellar distributions (contours) on the bottom left and right have
projected half-light radii of $0.184$ and $0.141$ length units,
respectively.}
\label{fig05}
\end{figure*}

In broad outline, the mergers resulting from these twelve encounters
are similar to those described in earlier studies (Negroponte \& White
1983; Barnes \& Hernquist 1991, 1996).  Table~\ref{tab:remnants} lists
merger times $t_{\rm merger}$ for all twelve encounters.  Gas usually
has little effect on the overall dynamics of orbit decay and merging;
of the four 1:1~C encounters, three merge within times $\Delta t_{\rm
merger} \simeq \pm 0.2$ of the corresponding stellar-dynamical
versions (Barnes 1998).  (The exception is RET~1:1~C, where the
mechanical friction of the gas hastens the merger by $\Delta t_{\rm
merger} \simeq 0.5$.)  As mentioned in the introduction, gas which
suffers strong shocks tends to fall into galactic nuclei before and
during the merger, where it creates central concentrations of highly
dissipated gas.  On the other hand, the gas which is {\it not\/}
involved in strong shocks retains a large fraction of its initial
angular momentum, and is available to build extended disks like those
shown in Figure~\ref{fig05}.

\begin{table}
\caption{Merger remnants.  Here $t_{\rm end}$ is the simulated time at
the end of the calculation, and $t_{\rm merger}$ is when the galaxies
merge.  The last five columns list gas percentages in remnant nuclei
({\bf N\/}), bars ({\bf B\/}), disks ({\bf D\/}), loops ({\bf L}), and
tails ({\bf T}).  These are rounded to the nearest percent, and do not
include unbound gas in the tails.  The {\bf B\/} entry for DIR 1:1 C
records the mass of the extra-nuclear ring in this remnant.}
\label{tab:remnants}
\begin{tabular}{@{}llrrrrrrr@{}}
{\bf Geom.} & {\bf Orbit} &  $t_{\rm end}$ & $t_{\rm merger}$ &
{\bf N} & {\bf B} & {\bf D} & {\bf L} & {\bf T} \\
\\
DIR & 1:1 C &   6.0 &  3.9 &  72 &   1 &  10 &     &  13 \\
    & 1:1 D &  10.0 &  8.5 &  85 &   2 &   2 &     &   8 \\
    & 3:1   &   6.0 &  4.3 &  79 &   1 &   8 &     &  10 \\
\\
RET & 1:1 C &   6.0 &  4.3 &  91 &     &   3 &   3 &   3 \\
    & 1:1 D &  11.5 &  9.5 &  93 &     &   5 &     &   2 \\
    & 3:1   &   6.0 &  5.5 &  89 &     &  11 &     &     \\
\\
POL & 1:1 C &   8.0 &  4.5 &  65 &     &  27 &   3 &   5 \\
    & 1:1 D &  12.0 &  9.8 &  66 &     &  21 &   4 &   8 \\
    & 3:1   &   8.0 &  6.3 &  52 &  10 &  22 &  14 &   2 \\
\\
INC & 1:1 C &   8.0 &  4.4 &  76 &     &  18 &     &   4 \\
    & 1:1 D &  12.0 &  8.1 &  42 &     &  51 &   1 &   5 \\
    & 3:1   &   8.0 &  6.4 &  75 &  10 &  11 &     &   4 \\
\end{tabular}
\end{table}

Ideally, simulations would be run well past $t_{\rm merger}$.  But as
already noted, the later stages of these calculations are quite slow.
Practical considerations forced me to stop at the times $t_{\rm end}$
listed in Table~\ref{tab:remnants}.  Most remnants had $t_{\rm end} -
t_{\rm merger} \simeq 2 \pm 1$ time units to relax; scaling the
initial disks to the Milky Way, the galaxies in these experiments
merged $\sim 10^8$ to~$10^9 {\rm\,yr}$ before the end of the
calculations.  These remnants may thus be compared to late-stage
Toomre-sequence (Toomre \& Toomre 1972; Toomre 1977) galaxies or to
IR-luminous merger remnants (Sanders \& Mirabel 1996).

\subsection{Structure}

\begin{figure*}
\begin{center}
\epsfig{figure=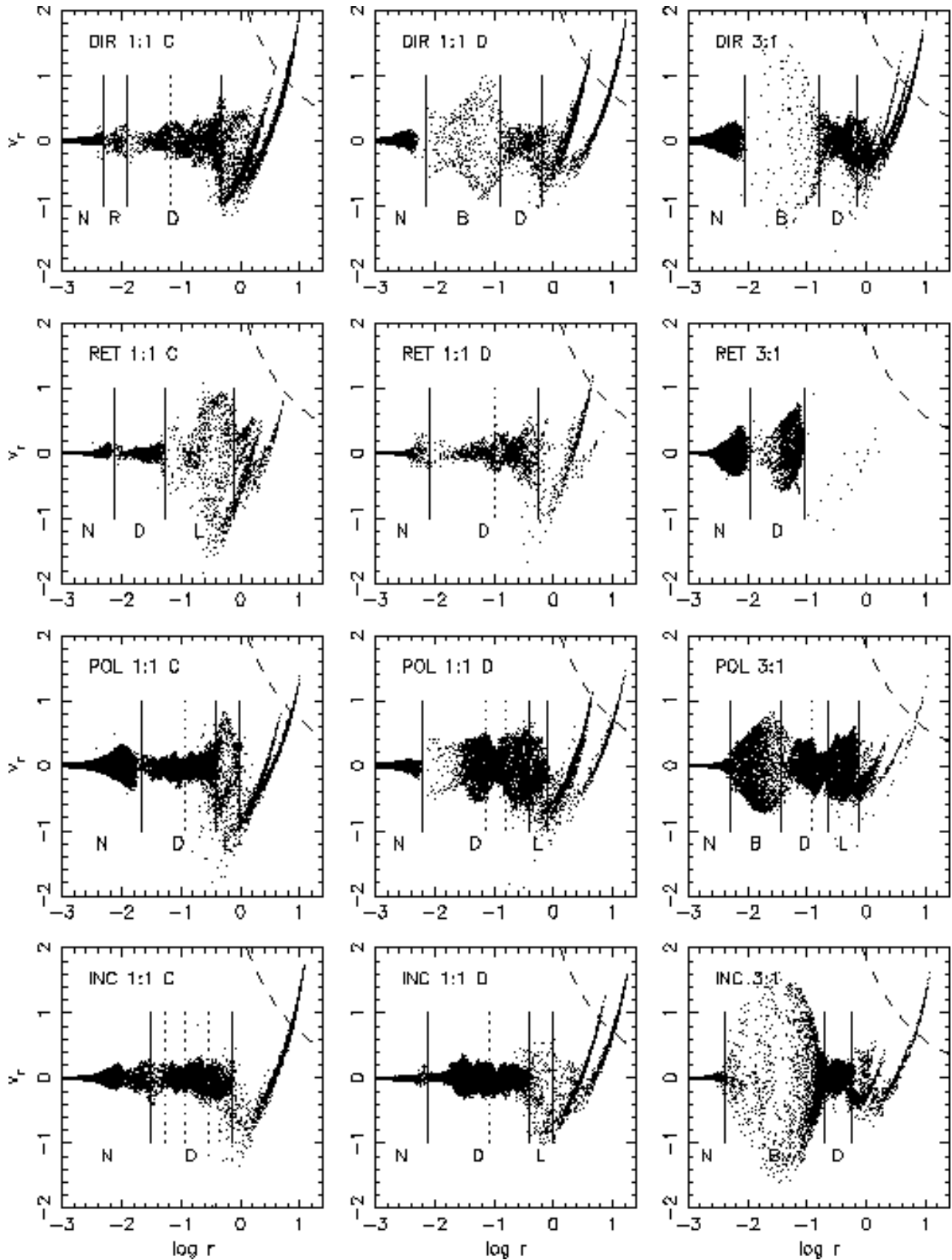,height=8.5in}
\end{center}
\caption{Radial velocities and distribution of gas in all merger
remnants.  In each plot, the horizontal axis is $\log r$, while the
vertical axis is $v_r$; the dashed curves show the escape velocity
$v_{\rm esc}$.  Solid vertical lines indicate divisions between major
components of the gas, labeled ``N'' (nuclei), ``R'' (rings), ``B''
(bars), ``D'' (disks), or ``L'' (loops).  Dotted vertical lines
roughly delimit warps in disks.}
\label{fig06}
\end{figure*}

Figure~\ref{fig06} summarizes the distribution and kinematics of gas
at the end of each simulation.  These plots show logarithmic radii
$\log r$ vs.\ radial velocities $v_r$ of gas particles.  The
distributions are roughly {\it symmetric\/} with respect to the
horizontal axis $v_r = 0$ at small radii, but distinctly {\it
asymmetric\/} at larger radii.  The narrow diagonal streaks seen at
large radii represent tails extracted from the disks before they
merged.  Almost every pericentric passage yields a pair of tails, but
the amount of tail material and its subsequent evolution depend on the
geometry of the passage.  Thus at the instants shown in
Figure~\ref{fig06} the remnant produced by RET~3:1 has almost no
gaseous tail material, while remnant DIR~3:1 still shows {\it three\/}
distinct tails (and the remains of a fourth), each originating from a
different disk at a different passage.  The dashed curves in the upper
right of each plot represent the nominal escape velocity $v_{\rm esc}
\equiv \sqrt{2 G (M_1 \! + \!  M_2) / r}$.  Some tail material has
enough velocity to escape, but almost all falls back eventually.  Tail
material with $v_r < 0$ has passed apogalacticon and is now falling
back; as this material reenters the remnant, the gas encounters shocks
and begins evolving toward closed orbits.

If the infalling gas has relatively little angular momentum it may
plunge deep into the remnant and directly assimilate into a disk.  In
many cases, however, the tail gas has so much angular momentum that it
initially side-swipes the central disk and subsequently follows
eccentric, looping paths before falling back yet again; a vivid
example is provided on the right in Figure~\ref{fig05}.  The gas on
these looping orbits populates transition zones between asymmetric and
symmetric $v_r$ distributions in Figure~\ref{fig06}.  These zones are
labeled by the letter ``L''; the various POLar encounters all produce
clear examples of this behavior.  A given gas particle may settle into
a closed orbit after executing a few loops, but the loop structures
themselves are replenished by continued infall.

At radii where the gas distribution is nearly symmetric with respect
to $v_r = 0$ lie the gas disks formed by this infall process.  Gas on
nearly-circular orbits has $v_r \simeq 0$, while gas on elongated
orbits has a large $v_r$ dispersion.  Thus in Figure~\ref{fig06} the
vertical spread of the points at a given radius gives information on
the shapes of gas orbits.  For example, remnants such as POL~1:1~C,
INC~1:1~C, and INC~1:1~D have large amounts of gas with fairly narrow
$v_r$ distributions; this gas lies in rotating disks, each labeled
``D'' in the plots.  At smaller radii, remnants DIR~1:1~D, DIR~3:1,
POL~3:1, and INC~3:1 have zones, labeled ``B'' in the plots, with
large, symmetric spreads in $v_r$.  These arise where central triaxial
or bar-like structures dominate the potential and the gas consequently
moves on elongated orbits.

All twelve remnants have nuclear gas clouds characterized by very high
densities, often of order $\rho \sim 10^6$ model units.  These nuclei,
which are labeled ``N'' in Figure~\ref{fig06}, usually have small
radial velocity dispersions; most are supported by a combination of
pressure and circular motion.  But remnants DIR~3:1, RET~3:1,
POL~1:1~C, and INC~1:1~C are different; in these cases the nuclear gas
lies in rotating elliptical disks.  These nuclei are barely resolved
in the present simulations, so their detailed dynamics are not
well-determined.  However, the total masses and angular momenta of
nuclear clouds should be more reliable since these are determined by
events on somewhat larger scales.  Several of these nuclear gas disks
are kinematically decoupled with respect to the remnants they inhabit.

The gas fractions in the nuclei, bars, disks, loops, and tails of all
remnants are listed in Table~\ref{tab:remnants}.  Here and in what
follows, each of these components is simply defined as {\it all\/} gas
within an appropriate range of radii measured from the center of each
remnant.  These radii often correspond to obvious features in
Figure~\ref{fig06}; for example, the ``D'' (disk) component in remnant
DIR~1:1~C is delimited by the abrupt end of the infalling tails at
$\log r \simeq -0.33$ on the one hand, and by a marked change in the
density of points at $\log r \simeq -1.92$ on the other hand.  In a
few cases the divisions between components are less obvious; in
particular, it's hard to define the boundaries between disks and outer
loops from the data in Figure~\ref{fig06}.  To fix these boundaries
more accurately, I used an interactive viewing program to inspect the
gas distribution in three dimensions; on such a display, as in the
right-hand side of Figure~\ref{fig05}, the distinction between nearly
circular disks and elongated loops is unmistakable.

As Table~\ref{tab:remnants} shows, the initial orientation of the
colliding disks has a definite effect on the distribution of gas in
the remnants.  Encounters in which {\it both\/} disks are tilted with
respect to the orbital plane yield remnants with massive disks, while
encounters in which a disk lies in the orbital plane drive more gas to
the nuclear regions.  Nuclei contain an average of $\sim 85$ percent
of the gas in the remnants of DIRect and RETrograde encounters; the
most massive nuclei, found in the RETrograde encounters, are produced
by violent hydrodynamic forces as illustrated in Figure~\ref{fig01}.
In contrast, only $\sim 63$ percent of the gas lies in the nuclei of
the POLar and INClined remnants; the balance is found in the extended
bars, disks, and loops surrounding these nuclei.

\begin{figure*}
\begin{center}
\epsfig{figure=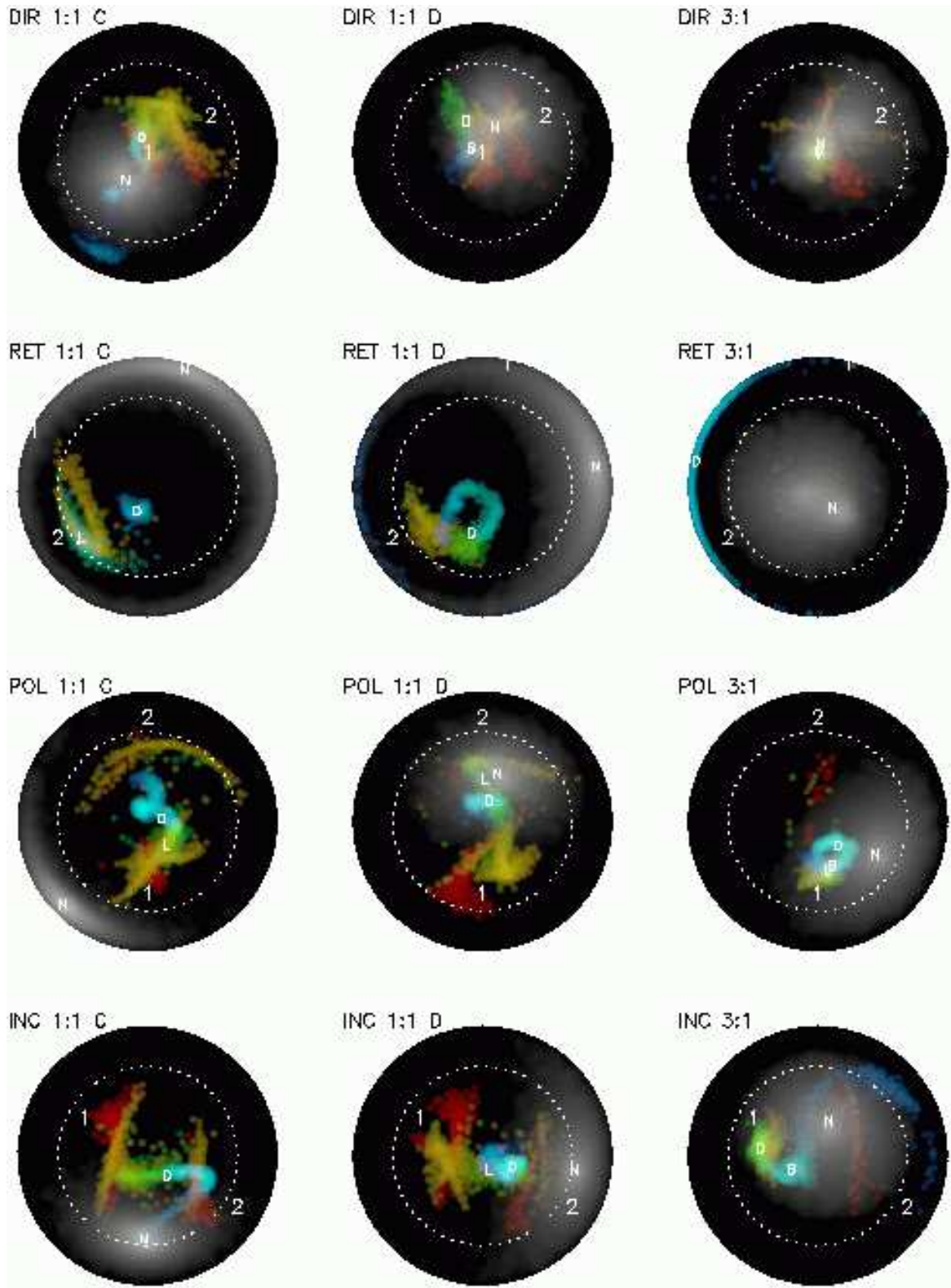,height=8.25in}
\end{center}
\caption{Rotation of gas components in merger remnants.  In each plot,
the sphere of all spin directions has been mapped into a circle whose
center coincides with the initial orbital angular momentum of the
galaxies; the dotted circle is $90^\circ$ from the orbital axis.
Images show angular momenta directions of gas particles; nuclei are
rendered in grey, while for other components colors represent radii,
progressing from blue for small $r$ to red for large $r$.  Numbers
``1'' and ``2'' indicate the initial spin axes of the galaxies, while
letters show spin axes of components defined in Figure~\ref{fig06}.}
\label{fig07}
\end{figure*}

Figure~\ref{fig07} displays rotation directions for the gas in the
remnants.  In each plot, an area-preserving map is used to transform
the unit sphere ${\bf S}$ of all possible spin directions onto the
unit circle ${\bf C}$; the direction of the initial orbit's angular
momentum vector ${\bf J}_{\rm orbit}$ always maps to the origin of
${\bf C}$.  Any area-preserving map from the sphere to the plane must
create some distortion; here, the direction exactly opposite to ${\bf
J}_{\rm orbit}$ maps to the {\it entire\/} edge of the circle, and
nearby regions are stretched into elongated crescents.  However, the
area-preserving property of this map implies that an isotropic
ensemble of directions is mapped to a uniform distribution within
${\bf C}$.  The images represent the angular momenta directions of gas
particles; each particle's angular momentum vector, measured with
respect to the center of the remnant as defined by the peak gas
density, was projected onto ${\bf S}$ and thence mapped to ${\bf C}$,
and the resulting distributions were smoothed with an iterated boxcar.
In these images, color is used to indicate radii, with grey for the
nuclei, blue for small radii, green and yellow for intermediate radii,
and red for large radii.

As Figure~\ref{fig07} shows, the remnants have diverse kinematics.
Many of the structures seen here can be identified with components
delineated in Figure~\ref{fig06}.  Gas in the nuclei, shown in grey,
produces the broadest distributions, but each nucleus has a
well-defined net spin direction which is marked by a ``N'' in these
plots.  Likewise, the ``B'', ``D'', and ``L'' symbols mark the net
spin directions of the bars, disks, and loops; these coincide with the
angular momenta of gas particles, shown in blue and green.  On the
other hand, the spin vectors of the {\it initial\/} disks, labeled
``1'' and ``2'', generally don't correspond to favored spin directions
in the remnants, but often appear associated with material in tidal
tails, here shown in yellow and red.

A few remnants display fairly simple structures.  For example,
remnants of DIRect encounters tend to have nuclei and disks with
reasonably well-aligned rotation, basically because the spins and
orbital motion of the initial encounter reinforce each another.  But
in many cases the gas has some kind of kinematic misalignment.  All
remnants of RETrograde passages have disjoint angular momenta
distributions, with counter-rotating nuclei and disks.  This is not
surprising given the rather contrived initial conditions of these
mergers; the distribution of gas between direct and retrograde
rotation reflects the competition of spin and orbital angular momenta
in these encounters.  Several other remnants have gas nuclei with
dramatic kinematic misalignments: remnant POL~1:1~C has a
counter-rotating nucleus, while remnants INC~1:1~C and~D have nuclei
which rotate about axes roughly perpendicular to the axes of their
disks.

\begin{figure*}
\begin{center}
\epsfig{figure=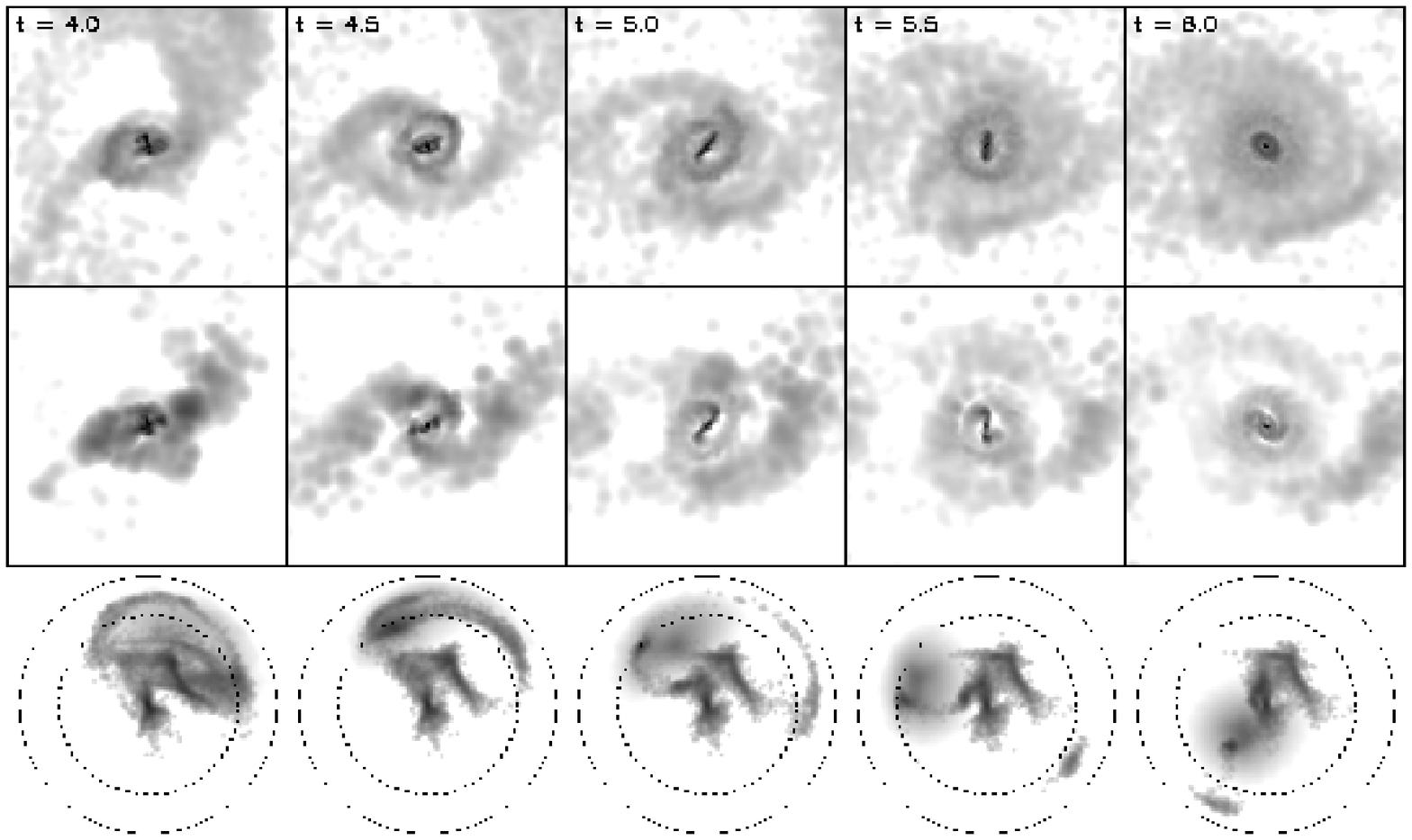,height=\textwidth,angle=-90}
\end{center}
\caption{Evolution of gas in remnant DIR~1:1~C.  The top and middle
rows view the gas distribution along the encounter's original orbital
axis; each frame is $1 \times 1$ length units.  In the top row all gas
particles are weighted equally, while in the middle row each particle
is weighted by the local dissipation rate $\dot{u}$.  In the bottom
row each particle's orbital angular momentum is mapped onto a circle;
between times $t = 4.5$ and $t = 6$, the nucleus and inner disk
precess counter-clockwise from 11 o'clock to 7 o'clock, while the
extra-nuclear ring precesses clockwise from 2 o'clock to 7 o'clock.}
\label{fig08}
\end{figure*}

As noted above, many remnant disks have warps.  A disk which warps
progressively along a single direction appears as a roughly linear
feature with a smooth color gradient in Figure~\ref{fig07}, as in the
plot for remnant INC~1:1~C.  In other disks the warp's direction
changes with radius, perhaps as a result of differential precession;
such warps produce curved forms as in the plot for remnant POL~1:1~C,
or even circular features like the one in RET~1:1~D.  Most of these
disks are fairly flat at small radii, as if the inner disks are locked
into a single plane by self-gravity.  Warps in the outer disks of
these remnants may be excited by the return of tail material with
misaligned angular momentum.  In Figure~\ref{fig07} these remnants
exhibit fairly smooth transitions from the disks (blue or green) to
the tails (yellow or red); thus the angular momentum of the gas
changes in a continuous manner with radius.  The tails contain large
amounts of angular momentum; as this material falls back it could
exert strong torques, creating long-lived warps in remnant disks.

Remnant DIR~1:1~C provides the example of a precessing inner disk
shown in Figure~\ref{fig08}.  This disk, bounded by the dotted line at
$\log r \simeq -1.2$ in the upper-left panel of Figure~\ref{fig06},
begins forming as soon as the nuclei of its progenitors have merged
and the gravitational potential has settled down.  By $t = 4.5$ it's
seen roughly edge-on extending from eight to two o'clock; it has
precessed counter-clockwise by $\sim 90^\circ$ by $t = 6$.  During
this time the inner disk remains relatively well-aligned with the
nucleus.  But between the nucleus and the inner disk lies a ring of
material, inclined by $\sim 130^\circ$, which precesses {\it
clockwise\/} by $\sim 120^\circ$ between $t = 4.5$ and~$6$.  If the
nucleus and inner disk are actually locked in alignment, they must be
coupled gravitationally, since the ring lies between them.


To rough approximation, both the inner disk and extra-nuclear ring in
this remnant behave like inclined rotators precessing in an oblate
potential.  But the middle row in Figure~\ref{fig08} provides evidence
that viscous forces also influence the dynamics of the inner disk.
These images show the gas particles weighted by their dissipation rate
$\dot{u}$; the inner disk is quite prominent, as is the gas just
beyond the inner disk.  The latter exhibits a spiral dissipation
pattern, implying that angular momentum is being transported radially.
One consequence of this viscous coupling may be the gradual decrease
in the inclination of the inner disk.

The inner disk of DIR~1:1~C acquired its tilt in a fairly
straightforward manner.  Some $74$ percent of the gas particles in
this component came from the $i = 71^\circ$ galaxy, and the angular
momentum of these particles determines the initial orientation of the
disk.  Some misaligned components in other remnants may be explained
in a similar way; in these cases the misaligned material comes
preferentially from one progenitor or the other.  But the relationship
between the originating disk and the final spin direction is not
always simple -- gravitational and hydrodynamical torques play a large
role in determining the final rotation direction of the gas.

\subsection{Infall}

The simulations cover only the early evolution of these merger
remnants; they include the violent dissipative events which form the
inner parts of gas disks but end shortly thereafter.  On longer
time-scales, much of the gas lingering in tidal tails will fall back
(Hibbard \& Mihos 1995), evolve through loop-like structures, and
finally seek out closed orbits.  It's computationally expensive to
follow this process much past the stage shown here, but counting the
bound tail gas, it appears that the DIRect and RETrograde remnants
have enough material to form disks containing up to $25$ percent of
the total gas, while the POLar and INClined mergers can form disks
containing $25$ to nearly~$60$ percent.

The time-scale for this infall is easily estimated in the limit where
the tail gas is just barely bound to the remnant.  In this limit the
tail material moves on nearly radial orbits in an approximately
Keplerian potential.  Suppose that a tail is formed during a
pericentric passage at time $t_{\rm peri}$.  A parcel of tail material
with specific binding energy $E < 0$ reaches apogalacticon at radius
$r_{\rm apo} \simeq G (M_1 \!  + \!  M_2) / |E|$, and falls back at
time
\begin{equation}
  t_{\rm ret} \simeq t_{\rm peri} +
    \frac{\pi G (M_1 \! + \! M_2)}{\sqrt{2} \, |E|^{3/2}} \, .
  \label{eq:tret}
\end{equation}
In this model there's a one-to-one relationship between the binding
energy distribution of the tail material and the rate at which this
material is reaccreted.  Let $d m/d E$ be the differential
distribution of binding energy of tail gas; then the reaccretion rate
is
\begin{equation}
  \frac{d m}{d t} = \frac{d m}{d E} \, \frac{d E}{d t_{\rm ret}}
    \propto \frac{d m}{d E} \, (t - t_{\rm peri})^{-5/3} \, .
\end{equation}
As Figure~\ref{fig06} shows, some tails are entirely bound; these fall
back completely after a finite time.  On the other hand, a tail which
extends to $E > 0$ generally has a distribution $d m/d E$ which is
approximately constant for $E \simeq 0$.  This follows because the
tidal interaction has no way to ``pick out'' the escape energy; thus
$d m/d E$ will be nonzero and continuous near $E = 0$ (e.g.~White
1987; Jaffe 1987).  Thus if some tail gas has enough energy to escape
then the reaccretion rate $d m/d t \propto t^{-5/3}$ as $t \to
\infty$.

\begin{figure}
\begin{center}
\epsfig{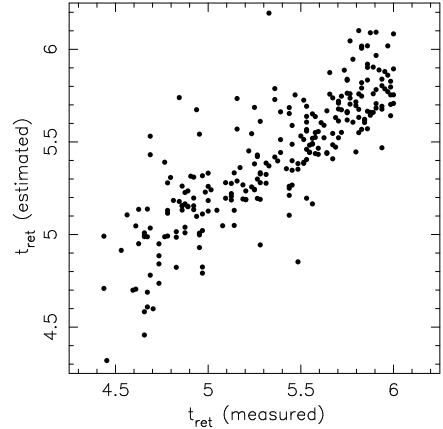}
\end{center}
\caption{Infall times for stellar particles in the first tails of
encounter DIR~1:1~C.}
\label{fig09}
\end{figure}

\begin{figure}
\begin{center}
\epsfig{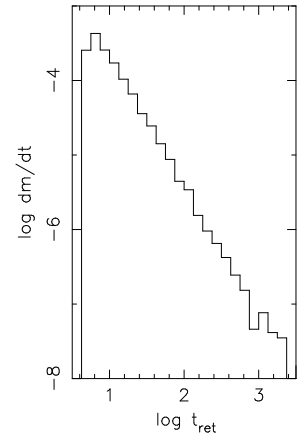}
\end{center}
\caption{Infall rate for gas from the tails produced in the first
passage of encounter DIR~1:1~C.}
\label{fig10}
\end{figure}

This model gives a crude but serviceable description of the infall of
tail material in remnant DIR~1:1~C.  Figure~\ref{fig09} shows infall
times for particles from the tails created at the first passage
($t_{\rm peri} = 1$); the horizontal axis is the measured time of
infall, while the vertical axis is the time predicted by
(\ref{eq:tret}) using specific binding energies determined at time $t
= 4$.  This comparison was made using collisionless particles from the
{\it stellar\/} disks: gas is deflected by shocks, so its exact
instant of perigalacticon is harder to define.  But both gas and stars
travel on nearly ballistic trajectories while in tails, so the rough
agreement seen here should also hold for the gas.  The scatter between
the predicted and measured infall times probably arises because actual
orbits are not radial and the true potential is not Keplerian.  At
later times the simple analytic model should grow more accurate since
late-returning material falls back from ever-larger distances.
Figure~\ref{fig10} shows the infall rate for gas from the
first-passage tails of DIR~1:1~C, predicted using binding energies
measured at $t = 4$.  Because of the way this sample of tail particles
was defined, the infall rate is depressed for $\log t \la 0.7$.  At
later times, however, the infall rate evolves almost precisely as
$t^{-5/3}$.  This is consistent with the distribution of binding
energies $d m/d E$ of gas in these tails, which varies by less than
$\pm 20$ percent for $-1.25 < E < 0.125$.

\begin{figure}
\begin{center}
\epsfig{figure=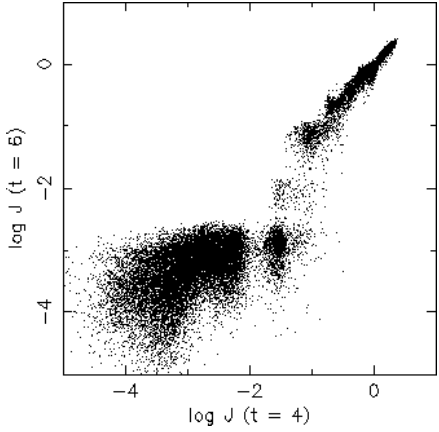,width=2.25in}
\end{center}
\caption{Specific angular momenta of gas particles at $t \simeq t_{\rm
merger}$ (horizontal axis) and $t = t_{\rm end}$ (vertical axis) for
remnant DIR~1:1~C.}
\label{fig11}
\end{figure}

If the infall time of tail gas is basically determined by its specific
binding energy $E$, its evolution once it falls back into the remnant
depends largely on its specific angular momentum $J$.  Studies of disk
formation (e.g.~Mestel 1963; Fall \& Efstathiou 1980) often assume
that each parcel of disk material conserves angular momentum during
proto-galactic collapse.  This can be tested in the simulations;
Figure~\ref{fig11} compares specific angular momenta of gas particles
at $t = 4 \simeq t_{\rm merger}$ and $t = 6 = t_{\rm end}$ in remnant
DIR~1:1~C.  This plot shows two main components: a broad spray of
points in the lower left, and a more or less diagonal distribution
extending to the upper right.  The former is the nucleus, while the
latter include the extended disk and tidal tails.  The nucleus
captures some gas between $t_{\rm merger}$ and $t_{\rm end}$, but most
of the gas which has enough angular momentum at the time of the merger
-- say $\log J \ga -1.4$ -- remains outside the nucleus and
approximately conserves the {\it magnitude\/} of its specific angular
momentum.

Detailed conservation of angular momentum can be used to predict the
radial distribution of the gas at later times.  For simplicity, assume
that the gas eventually finds circular orbits in the potential
$\Phi(r)$ of a spherical mass distribution.  Then a gas particle's
specific angular momentum $J$ and orbital radius $r$ are related by
\begin{equation}
  J = \sqrt{r^3 \, d \Phi / d r} \, .
  \label{eq:jcirc}
\end{equation}
Let $d m/d J$ be the differential angular momentum distribution of the
gas; then the predicted radial distribution in the final disk is
\begin{equation}
  \frac{d m}{d r} = \frac{d m}{d J} \, \frac{d J}{d r} \, .
\end{equation}
To predict how the disk's radial profile evolves as infall proceeds,
(\ref{eq:tret}) may be used to restrict $d m/d J$ to only the gas
which has fallen in before a given time.

\begin{figure}
\begin{center}
\epsfig{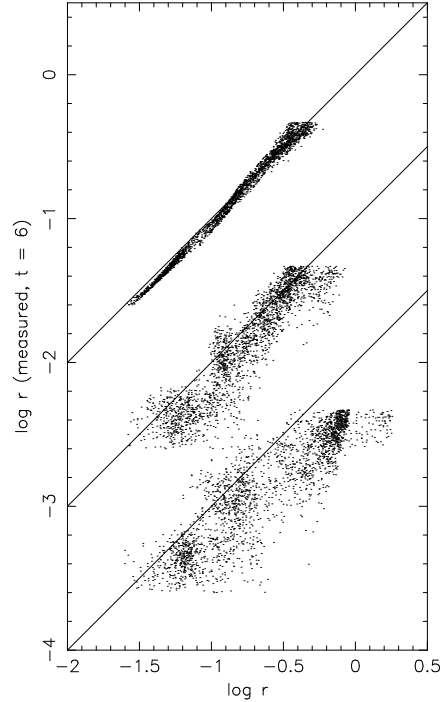}
\end{center}
\caption{Predicted and measured distributions of gas from the disk of
remnant DIR~1:1~C.  The vertical axis shows the radius of each gas
particle measured at time $t = 6$; successive plots are shifted by
$-1$ in $\log r$.  Top: the horizontal axis is the radius {\it
calculated\/} using $J$ measured at $t = 6$.  Middle: the radius {\it
predicted\/} using $J$ measured at $t = 4$.  Bottom: the radius {\it
measured\/} at $t = 4$.}
\label{fig12}
\end{figure}

Figure~\ref{fig12} illustrates the prediction of radial coordinates
for gas particles in remnant DIR~1:1~C using (\ref{eq:jcirc}) and the
model potential in Appendix~A.  Three different scatter-plots are
presented; in every one the vertical axis shows radii of gas particles
which lie in the range $-1.6 < \log r < -0.33$ at time $t = 6$.  In
the top plot the horizontal axis shows radii for these particles
calculated from (\ref{eq:jcirc}) using angular momenta $J$ measured at
$t = 6$; the points fall almost precisely on the diagonal line,
confirming that the gas particles are on nearly circular orbits and
that the mass model is adequate.  In the middle plot the same
equations are used, but the angular momenta $J$ are measured at $t =
4$, which is just after the merger.  If individual gas particles
conserved $J$ precisely, this plot would be identical to the top one;
the actual distribution is considerably broader, but still more or
less straddles the diagonal line.  Finally, in the bottom plot the
horizontal axis shows the radii measured just after the merger at $t =
4$; this distribution is broader still and distinctly offset from the
diagonal line, showing that significant infall has taken place between
$t = 4$ and $t = 6$.

In sum, Figure~\ref{fig12} represents a qualified success for this
simple model of disk formation by infall from tails.  Given a model
potential $\Phi(r)$, the radial coordinate of a particle of disk gas
can be calculated with some accuracy if its {\it final\/} angular
momentum is known.  If the angular momentum at an earlier time is
given instead, it's still possible to roughly predict the final
radius.  Such predictions show some scatter, but they're better than
simply taking radii at the moment of merger as estimates of final
radii.  The scatter could be reduced if the torques acting on the gas
between $t_{\rm merger}$ and $t_{\rm end}$ were taken into account --
but this is beyond the scope of the present model.

\begin{figure}
\begin{center}
\epsfig{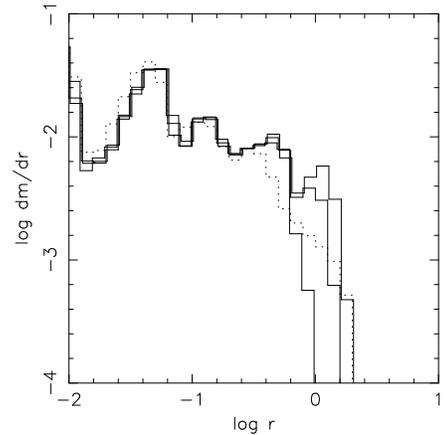}
\end{center}
\caption{Radial distributions of gas.  Dotted curve is measured
distribution of gas with $E < -1$ at $t = 6$.  Solid curves are {\it
predicted\/} final distributions of gas; from left to right, these
curves show gas with $E < -1$, $E < -0.5$, and $E < 0$, respectively.}
\label{fig13}
\end{figure}

Figure~\ref{fig13} illustrates the use of (\ref{eq:jcirc}) to predict
the future growth of the disk in remnant DIR~1:1~C.  Here the dotted
curve shows the actual radial distribution of all gas particles with
$E < -1$ at $t = 6$; this subset includes the disk and some of the
infalling tail material.  The three solid curves show {\it
predicted\/} radial distributions of gas particles with $E < -1$, $E <
-0.5$, and $E < 0$.  At radii $\log r \la -0.4$ all particles are on
essentially circular orbits and all four curves agree.  At larger
radii the actual distribution is more extended than the predicted
distribution for $E < -1$ since outlying gas has not yet had a chance
to settle into circular orbits.  The distribution predicted for $E <
-1$ has a fairly sharp break at $\log r \simeq -0.2$ and essentially
vanishes at $\log r \simeq 0$.  This reflects the fact that there is a
fairly well-defined ceiling on the angular momenta of gas particles
with $E < -1$; binding energies and angular momenta in tails are
correlated, with the less tightly bound tail material having more
angular momentum.  Gas with $-1 < E < -0.5$ has more angular momentum
than gas with $E < -1$, and therefore settles at larger radii; gas
with $-0.5 < E < 0$ settles at larger radii still.  Thus the disk will
grow from the inside-out, with reaccretion constantly depositing gas
at the outer edge of the disk.  This can be seen in
Figure~\ref{fig08}; the top row of images show the radius of the disk
increasing with time, and the middle row shows that the periphery of
the disk has a high rate of dissipation as infalling material collides
with disk gas.

\section{DISCUSSION}

Gas disks like those described here were briefly reported in earlier
simulations of dissipative mergers between disk galaxies (Hernquist \&
Barnes 1991; Barnes \& Hernquist 1996).  These studies used the same
general methodology as the present work, but differed in several
specifics.  First of all, in earlier work the thermal evolution of the
gas was explicitly computed, including radiative cooling with a
cut-off at $T = 10^4 {\rm\,K}$.  This had little consequence for the
gas in regions of moderate to high density where cooling times are
short; however, the gas tails in the present experiments are broader
and smoother since they are not cooled by adiabatic expansion.
Second, the earlier studies used King (1966) distributions for the
bulges and halos of the initial galaxy models.  As a result, these
models may have been more susceptible to bar instabilities (Mihos \&
Hernquist 1996), and their relative orbits decayed appreciably faster
(Barnes 1998).  Third, the earlier simulations used the TREESPH code
(Hernquist \& Katz 1989); the code used in the new simulations borrows
several features from TREESPH, but it adopts a different smoothing
prescription for gravitational forces, a more efficient and robust
tree algorithm, a fixed number of neighbors in SPH summations, and a
different time-step algorithm.  Despite using somewhat different
models and codes, these studies all yield broadly similar results; the
process of disk formation appears to be insensitive to such details.

Simulations of polar ring formation in galaxy mergers have produced
remnants with very large rings or annuli (Bekki 1998).  These rings
arise from a relatively restricted set of initial conditions which
don't overlap with the ones considered here.

Disk formation is also seen in simulations of hierarchical clustering
which include gas as well as cold dark matter (e.g.~Evrard, Summers, \&
Davis 1994; Navarro \& Steinmetz 1997; Dom\'{\i}nguez-Tenreiro,
Tissera, \& S\'aiz 1998).  These disks are generally identified with
{\it spiral\/} galaxies, and they form largely from pristine gas which
was not previously part of another galaxy.  In common with the disks
seen in the present experiments, disks formed by hierarchical
clustering generally grow from the inside out.

\subsection{Origins of counter-rotation}

A rather striking result from this study is the large number of
remnants with counter-rotating or otherwise misaligned nuclei.  In
some cases, counter-rotation is easily explained; for example, it's no
great surprise that the RETrograde encounters all produce remnants
with counter-rotating components.  Likewise, some misaligned
components appear to contain gas originating almost exclusively from
one of the two progenitors, and it's likely that such ``unblended''
structures retain some memory of the orientation of their parent
disks.  But the origin of the misalignment is not always so evident.

The nuclear disk in remnant POL~1:1~C is a case in point.  In terms of
overall structure this remnant is very similar to one previously
described (Hernquist \& Barnes 1991); in both a fairly massive nuclear
gas disk counter-rotates with respect to an outer gas disk and stellar
component.  This resemblance may not seem surprising, since the same
initial disk geometry ($i_1 = 71^\circ$, $\omega_1 = 90^\circ$, $i_2 =
-109^\circ$, $\omega_2 = 90^\circ$) was used in both experiments.  But
the initial orbit used in the previous study was {\it wider\/}, with a
first pericenter at $r_{\rm peri} = 0.4$.  Of the encounters in the
present study, the one with initial conditions most closely matching
those used previously is POL~1:1~D, and POL~1:1~D does {\it not\/}
yield a counter-rotating nucleus.  It's not clear why this encounter
doesn't produce a counter-rotating nucleus like the one seen in the
earlier study, while it's close relative POL~1:1~C provides a very
good match.  One possibility is that the halos used in the present
study are a bit less effective at promoting orbital decay; as a
result, the merging nuclei arrive at their final encounter with
slightly more orbital angular momentum than they had in the earlier
experiment.

The mechanism which forms counter-rotating nucleus in objects like
remnant POL~1:1~C remains obscure.  The nucleus of this remnant is
already counter-rotating when it forms, and it's difficult to
determine when and how this material first acquired its rotation;
angular momentum can only be measured with respect to a center, and
before the nucleus forms the system has no unique center.  This
ambiguity in defining the center could actually be part of the puzzle
-- material co-rotating with respect to the center as defined in one
way may be counter-rotating with respect to the center defined in
another.  Large-scale hydrodynamic interactions seem to be important;
many of the counter-rotating or misaligned nuclei in the present
sample form when galaxy pairs arrive at their second passage still
dressed in extensive gas disks.  A better understanding of the roles
of gravitational and hydrodynamic torques in the formation of
counter-rotating nuclei may require calculations with higher spatial
resolution than available at present.

\subsection{Observational connections}

The results presented here invite comparisons with observations of gas
disks in merger remnants.  As mentioned in the introduction, NGC~7252
is an obvious example; the central disk of ionized and molecular gas
in this system is about the same size as the disks produced in these
experiments, and the peculiar kinematics of this disk may be due to a
strong warp of the kind seen in many of these simulations.  In
addition, H{\small I} velocities confirm that gas is falling back into
the remnant from the tidal tails (Hibbard et al.~1994).  NGC~7252 has
been modeled as the result of a close and fairly direct encounter
between two disk galaxies of comparable mass (Hibbard \& Mihos 1995).
This model nicely reproduced the morphology and kinematics of the
tails, but the calculations did not include a dissipative component
which could form a central disk.  It's worth repeating this
calculation with a combined N-body/SPH code; a single model
reproducing both the tails {\it and\/} the central disk would be a
fairly impressive accomplishment, and might help constrain the gas
content of NGC~7252's progenitors.

The peculiar elliptical galaxy NGC~3656 (Balcells 1997, Balcells et
al. 2001) may be a second example.  This galaxy contains shells and a
pair of faint tails which suggest a merger of two disk galaxies.  It
also has a star-forming dust lane which corresponds to an extended
disk of H{\small I}.  This disk is visibly warped, and the outer edge
of the disk is kinematically contiguous with gas at larger radii which
may be falling in from the tidal tails.  The estimated gas accretion
rate in NGC~3656 is an order of magnitude lower than in NGC~7252; this
may imply that NGC~3656 is at a later stage in its evolution than
NGC~7252.

NGC~5128, an elliptical galaxy with an active nucleus and an extended
disk of dust and gas, may represent an even later stage in the
evolution of gas-rich merger remnants.  Although the warped disk in
this system has been interpreted as the remains of an accreted
gas-rich satellite (e.g.~Malin, Quinn, \& Graham 1983), there are some
grounds to suspect that this galaxy is the result of a fairly major
merger (e.g.~Schweizer 1998).  These include the ripples visible in
deep optical images (Malin et al.~1983), the H{\small I} fragments
surrounding the galaxy (Schiminovich et al.~1994), and the misaligned
rotation revealed by kinematics of planetary nebulae (Hui et
al.~1995).

{\it Nuclear\/} gas disks and rings seem to be very common in merging
galaxies selected by infrared luminosity (e.g.~Downes \& Solomon
1998).  In remnants with single nuclei the gas tends to rotate more
rapidly than the stars, indicating some degree of kinematic decoupling
(Genzel et al. 2001), though it's not clear if the extreme decoupling
in remnant POL~1:1~C has yet been seen in the observations.  Disks in
systems which appear to still have {\it two\/} nuclei are not so
easily explained.  Molecular-line observations of Arp~220 and NGC~6240
have been interpreted in terms of a rotating disk located {\it
between\/} the nuclei (Scoville, Yun, \& Bryant 1997; Tacconi et
al.~1999; Tecza et al.~2000).  In the simulations, gas driven inward
before the galaxies merge always accumulates in disks or bars around
the individual nuclei; moreover, an incipient disk between the nuclei
would be torn apart by gravitational fields as the orbits of the
nuclei decay.  An inter-nuclear disk might be stable if it was more
massive than either nucleus (Tecza et al.~2000), but the amount of gas
required seems quite extravagant, and earlier reports of a peak in the
stellar velocity dispersion between the nuclei of NGC~6240 (Lester \&
Gaffney 1994; Doyon et al.~1994) are not supported by recent HST
observations (Gerssen et al.~2001).  On the other hand, the
interpretation of Arp~220's nuclei as bright spots in ``a warped
molecular gas disk'' (Eckart \& Downes 2001; Downes \& Solomon 1998)
seems entirely consistent with the present numerical results.

\subsection{Galaxy transformation}

Simulations indicate that extended gas disks can form in mergers of
spiral galaxies.  These disks contain between $20$ and~$60$ percent of
the total gas in the initial galaxies, and may extend to several times
the remnant half-light radii.  It's worth noting that the initial
galaxy models used in these experiments were not particularly
gas-rich, and that the gas started with the same distribution as the
disk stars.  In real spiral galaxies the atomic gas is more extended
than the stars; mergers of such systems should form remnants with even
larger and more massive gas disks.  Moreover, the gas content of disk
galaxies is generally expected to increase with redshift; remnants
with disks containing $\sim 10$ to~$15$ percent of their luminous mass
would very likely result by doubling the gas fraction in the initial
galaxy models.

These disks, if subsequently converted to stars, would be fairly hard
to detect photometrically (Rix \& White 1990) unless viewed from a
favorable orientation (e.g.~Scorza \& Bender 1990).  On the other
hand, disks could significantly influence the observed {\it
kinematics\/} of merger remnants.  Disks are kinematically cold, so
stars formed in such disks would add narrow features to the stellar
absorption-line spectra of early-type galaxies.  Moreover, disks
rotate faster than pressure-supported components, so these narrow
features will be systematically offset from the broader profiles due
to the rest of the galaxy.  The combined velocity profile of a disk
plus spheroid thus appears asymmetric, with a steep prograde wing and
a shallow retrograde wing (e.g.~Franx \& Illingworth 1988; Bender
1990; Rix \& White 1992; van der Marel \& Franx 1993).  Such profiles
actually appear to be fairly typical in elliptical galaxies with
measurable rotation (Bender, Saglia, \& Gerhard 1994).

The observed velocity profiles may be compared with those predicted
for remnants of purely dissipationless mergers between disk galaxies
(Bendo \& Barnes 2000, Naab \& Burkert 2001).  As a rule, simulated
equal-mass mergers produce remnants with the {\it opposite\/} velocity
profile asymmetry -- that is, they have shallow prograde wings and
steep retrograde wings.  For unequal-mass mergers the situation is not
so clear -- one study finds profiles similar to those produced in
equal-mass mergers, while the other reports cases which exhibit the
same sense of asymmetry seen in observations of early-type galaxies.
Naab \& Burkert (2001) suggested that the formation of extended disks
is {\it necessary\/} if mergers are to account for the observed
kinematics of elliptical galaxies, but worried that the gas might not
retain enough angular momentum to build extended disks.  Given the
contradictory results on unequal-mass mergers, it may be premature to
insist that all mergers must form disks.  On the other hand, the
present work indicates that disks of the requisite size can form quite
easily in both equal-mass and unequal-mass mergers.

If the above explanation for line profiles in elliptical galaxies is
correct, one might expect different spectral lines to show different
kinematic signatures.  The lines produced by the younger stars making
up a disk should be narrower than, and systematically offset from, the
lines produced by the older stars in the spheroid.  This might be
tested with high signal-to-noise spectra obtained using large optical
telescopes.

\section*{ACKNOWLEDGMENTS}

I thank the referee, A.~Burkert, for a constructive report, and
M.~Balcells, L.~Hernquist, J.~Hibbard, D.~Sanders, and F.~Schweizer
for valuable discussions.  This work was supported in part by STScI
grant GO-06430.03-95A.

\appendix

\section{REMNANT POTENTIAL}

A simple model of the potential may be constructed using a Plummer
model to represent the nucleus and a Hernquist model to represent the
rest of the remnant:
\begin{equation}
  \Phi(r) = - \frac{G M_{\rm nuc}}{(r^2 + a_{\rm nuc}^2)^{1/2}} -
                \frac{G M_{\rm rem}}{r + a_{\rm rem}} \, .
  \label{eq:massmodel}
\end{equation}
For remnant DIR~1:1~C, I took $M_{\rm nuc} = 0.034$ to be the mass
within $r < 0.005$, $a_{\rm nuc} = \epsilon = 0.0125$ to be the
``softened'' radius of the nucleus, $M_{\rm rem} = M_1 + M_2 - M_{\rm
nuc} = 2.466$ to be the mass of the rest of the remnant, and $a_{\rm
rem} = 0.2128$ to be the radius enclosing a quarter of the remnant's
mass exclusive of the nucleus.


\begin{thebibliography}{}

\bibitem[]{B97}
Balcells, M. 1997, ApJ 486, L87

\bibitem[]{BvGSdB01}
Balcells, M., van Gorkom, J.H., Sancisi, R., del Burgo, C. 2001,
astro-ph/0107165

\bibitem[]{B88}
Barnes, J.E. 1988, ApJ 331, 699

\bibitem[]{B92}
Barnes, J.E. 1992, ApJ 393, 484

\bibitem[]{BH91}
Barnes, J.E. \& Hernquist, L. 1991, ApJ 370, L65

\bibitem[]{BH96}
Barnes, J.E. \& Hernquist, L. 1996, ApJ 471, 115

\bibitem[]{B98}
Barnes, J.E. 1998, {\it Galaxies: Interactions and Induced Star
Formation\/}, eds. D. Friedli, L. Martinet, D. Pfenniger
(Springer-Verlag: Berlin), p.~275

\bibitem{BB97}
Bate, M.R. \& Burkert 1997, MNRAS 288, 1060

\bibitem{Be98}
Bekki, K. 1998, ApJ 499, 635

\bibitem[]{B90}
Bender, R. 1990, A\&A 229, 441

\bibitem[]{BSG94}
Bender, R., Saglia, R.P., \& Gerhard, O.E. 1994, MNRAS 269, 785

\bibitem[]{BB00}
Bendo, G.J. \& Barnes, J.E. 2000, MNRAS 316, 315

\bibitem[]{CDG90}
Combes, F, Dupraz, C., \& Gerin, M. 1990 {\it Dynamics and
Interactions of Galaxies\/}, ed. R. Wielen (Springer-Verlag: Berlin),
p.~205

\bibitem[]{CHSS93}
Condon, J.J., Helou, G., Sanders, D.B., \& Soifer, B.T. 1993, AJ 105,
1730

\bibitem[]{D93}
Dehnen, W. 1993, MNRAS 265, 250

\bibitem[]{DTS98}
Dom\'{\i}nguez-Tenreiro, R., Tissera, P.B., \& S\'aiz, A. 1998,
ApJ 508, L123

\bibitem[]{DS98}
Downes, D. \& Solomon, P.M. 1998, ApJ 507, 615

\bibitem{D+94}
Doyon, R., Wells, M., Wright, G.S., Joseph, R.D., Nadeau, D., \&
James, P.A. 1994, ApJ 437, 23

\bibitem{DCCK90}
Dupraz, C., Casoli, F., Combes, F., \& Kazes, I., A\&A 228, L5

\bibitem[]{ED01}
Eckart, A. \& Downes, D. 2001, ApJ 551, 730

\bibitem[]{ESD94}
Evrard, A.E., Summers, F.J., \& Davis, M. 1994, ApJ 422, 11

\bibitem[]{FE80}
Fall, S.M. \& Efstathiou, G. 1980, MNRAS 193, 189

\bibitem[]{FI88}
Franx, M. \& Illingworth, G.D. 1988, ApJ 327, L55

\bibitem[]{F70}
Freeman, K.C. 1970, ApJ 160, 811

\bibitem[]{GTRLM01}
Genzel, R., Tacconi, L.J., Rigopoulou, D., Lutz, D., \& Tecza,
M. 2001, astro-ph/0106032

\bibitem[]{GI97}
Gerritsen, J.P.E. \& Icke, V. 1997, A\&A 325, 972

\bibitem[]{GvdMAMHB01}
Gerssen, J., van der Marel, R.P., Axon, D., Mihos, C., Hernquist, L.,
\& Barnes, J.E. 2001, {\it The Central kpc of Starbursts and AGN\/},
ed. J.H. Knapen, J.E. Beckman, I. Shlosman, \& T.J. Mahoney (PASP: San
Francisco), in press

\bibitem[]{H89}
Hernquist, L. 1989, Nature 340, 687

\bibitem[]{H90}
Hernquist, L. 1990, ApJ 356, 359

\bibitem[]{H93}
Hernquist, L. 1993, ApJS 86, 389

\bibitem[]{HB91}
Hernquist, L. \& Barnes, J.E. 1991, Nature 354, 210

\bibitem[]{HK89}
Hernquist, L. \& Katz, N. 1989, ApJS 70, 419

\bibitem[]{H+94}
Hibbard, J.E., Guhathakurta, P., van Gorkom, J.H., Schweizer, F. 1994,
AJ 107, 67

\bibitem[]{HM95}
Hibbard, J.E. \& Mihos, J.C. 1995, AJ 110, 140

\bibitem[]{HVY00}
Hibbard, J.E., Vacca, W.D., \& Yun, M.S. 2000, AJ 119, 1130

\bibitem[]{HvdHBR01}
Himmard, J.E., van der Hulst, J.M., Barnes, J.E., \& Rich, R.M. 2001,
AJ 122, in press.

\bibitem[]{HFFD95}
Hui, X., Ford, H.C., Freeman, K.C., \& Dopita, M.A. 1995, ApJ 449, 592

\bibitem[]{J87}
Jaffe, W. 1987, {\it Structure and Dynamics of Elliptical Galaxies\/},
ed. T. de Zeeuw (D. Reidel: Dordrecht), p. 511

\bibitem[]{JW85}
Joseph, R.D. \& Wright, G.S. 1985, MNRAS 214, 87

\bibitem[]{K92}
Katz, N. 1992, ApJ 391, 502

\bibitem[]{K66}
King, I.R. 1966, AJ 71, 64

\bibitem[]{KS92}
Kormendy, J. \& Sanders, D.B. 1992, ApJ 390, L53

\bibitem[]{LT78}
Larson, R.B. \& Tinsley, B.M. 1978, ApJ 219, 46

\bibitem[]{LG94}
Lester, D.F. \& Gaffney, N.I. 1994, ApJ 431, L13

\bibitem[]{MQG83}
Malin, D.F., Quinn, P.J., \& Graham, J.A. 1983, ApJ 272, L5

\bibitem[]{M63}
Mestel, L. 1963, MNRAS 126, 553

\bibitem[]{M01}
Mihos, J.C. 2001, ApJ 550, 94

\bibitem[]{MH94}
Mihos, J.C. \& Hernquist, L. 1994, ApJ 437, 611

\bibitem[]{MH96}
Mihos, J.C. \& Hernquist, L. 1996, ApJ 464, 641

\bibitem[]{M92}
Monaghan, J.J. 1992, ARA\&A 30, 543

\bibitem[]{NB01}
Naab, T. \& Burkert, A. 2001, astro-ph/0103476

\bibitem[]{NS97}
Navarro, J.F. \& Steinmetz, M. 1997, ApJ 478, 13

\bibitem[]{NW83}
Negroponte, J. \& White, S.D.M. 1983, MNRAS 205, 1009

\bibitem[]{N88}
Noguchi, M. 1988, A\&A 203, 259

\bibitem[]{RW90}
Rix, H.-W. \& White, S.D.M. 1990, MNRAS 254, 389

\bibitem[]{RW92}
Rix, H.-W. \& White, S.D.M. 1992, ApJ 362, 52

\bibitem[]{SM96}
Sanders, D.B. \& Mirabel, I.F. 1996, ARA\&A 34, 749

\bibitem[]{SvGvdHK94}
Schiminovich, D., van Gorkom, J.H., van der Hulst, J.M., \& Kasow,
S. 1994, ApJ 423, L101

\bibitem[]{S82}
Schweizer, F. 1982, ApJ 252, 455

\bibitem[]{S98}
Schweizer, F. 1998, {\it Galaxies: Interactions and Induced Star
Formation\/}, eds. D. Friedli, L. Martinet, D. Pfenniger
(Springer-Verlag: Berlin), p.~105

\bibitem[]{SB90}
Scorza, C. \& Bender, R. 1990, A\&A 235, 49

\bibitem[]{SYB97}
Scoville, N.Z., Yun, M.S., \& Bryant, P.M. 1997, ApJ 484, 702

\bibitem[]{S42}
Spitzer, L. 1942, ApJ 95, 329

\bibitem[]{S00}
Springel, V. 2000, MNRAS 312, 859

\bibitem[]{TGTGDS99}
Tacconi, L.J., Genzel, R., Tecza, M., Gallimore, J.F., Downes, D., \&
Scoville, N.Z. 1999, ApJ 524, 732

\bibitem[]{TGTATGT00}
Tecza, M., Genzel, R., Tacconi, L.J., Anders, S., Tacconi-Garman,
L.E., \& Thatte, N. 2000, ApJ 537, 178

\bibitem[]{T64}
Toomre, A. 1964, ApJ 139, 1217

\bibitem[]{T77}
Toomre, A. 1977, {\it The Evolution of Galaxies and Stellar
Populations\/}, eds. B. Tinsley \& R. Larson (Yale University Obs.:
New Haven), p. 401

\bibitem[]{TT72}
Toomre, A. \& Toomre, J. 1972, ApJ 178, 623

\bibitem[]{TRBDFGKL94}
Tremaine, S., Richstone, D.O., Byun, Y.-I., Dressler, A., Faber, S.M.,
Grillmair, C., Kormendy, J., Lauer, T.R. 1994, AJ 107, 634

\bibitem[]{TKW98}
Tsuchiya, T., Korchagin, V., \& Wada, K. 1998, ApJ 505, 607

\bibitem{vdMF93}
van der Marel, R.P. \& Franx, M. 1993, ApJ 407, 525

\bibitem[]{W87}
White, S.D.M. 1987, {\it Structure and Dynamics of Elliptical
Galaxies\/}, ed. T. de Zeeuw (D. Reidel: Dordrecht), p. 339

\bibitem[]{WSLBR93}
Whitmore, B.C., Schweizer, F., Leitherer, C., Borne, K., \& Robert,
C. 1993, AJ 106, 1354

\end{thebibliography}
\end{document}